\documentclass[british]{scrartcl}
\usepackage[T1]{fontenc}
\usepackage[utf8]{inputenc}
\usepackage{geometry}
\usepackage{amsmath}
\usepackage{graphicx}

\makeatletter

\providecommand{\tabularnewline}{\\}

\graphicspath{ {Bilder/} }

\makeatother

\usepackage{babel}
\begin{document}

\title{Hard-Core Bosons in flat band systems above the critical density}

\author{Moritz Drescher and A. Mielke\thanks{Institut für Theoretische Physik, Universität Heidelberg, Philosophenweg
19, D-69121 Heidelberg, Germany}}
\maketitle
\begin{abstract}
We investigate the behaviour of hard-core bosons in one- and two-dimensional
flat band systems, the chequerboard and the kagomé lattice and one-dimensional
analogues thereof. The one dimensional systems have an exact local
reflection symmetry which allows for exact results. We show that above
the critical density an additional particle forms a pair with one
of the other bosons and that the pair is localised. In the two-dimensional
systems exact results are not available but variational results indicate
a similar physical behaviour.
\end{abstract}

\section{Introduction}

Tight binding models are important to study the properties of correlated
fermions or bosons. They describe interacting particles on a lattice.
The most popular and important model of that kind is the Hubbard model
\cite{Hubbard63,Kanamori63,Gutzwiller1963}, which was originally
formulated for fermions and is known to describe many different phenomena,
depending on the lattice structure and the interaction strength. The
bosonic Hubbard model was introduced by Fisher et al. \cite{Fisher1989}.
It is expected to show a similar rich phase diagram including a Mott
insulator and a superfluid phase. 

In the Hubbard model, the interaction of the particles is purely local
and the hopping is often a nearest neighbour hopping. The hopping
gives rise to bands. Strong correlations occur if the interaction
strength is large compared to the band width. 

Depending on the hopping matrix and on the structure of the lattice,
one or more bands may be completely flat. A flat band system is of
special interest if the physics of the system is mainly determined
by particles in the flat band. For fermions this happens if at low
temperatures the flat band is partially filled. For bosons the flat
band must be at the bottom of the single particle spectrum to have
a significant influence on the low temperature properties. In both
cases, the interaction, even if it is small, dominates the behaviour
of the system. A standard perturbational treatment of the interaction
fails. Even an arbitrarily small interaction is larger than the band
width and interesting correlation effects may occur. Flat band systems
have been studied for more than 25 years, starting with magnetic properties
of fermions in a flat band, see \cite{Lieb89,Mielke1991,Tasaki92,Mielke1993,Tasaki97b,Mielke1999}
and the references therein. Starting in 2002, spin systems have been
investigated in lattices with a flat band \cite{Schulenburg2002},
see also \cite{Derzkho2007} and the references therein. Approximately
seven years ago flat band systems were first studied experimentally
using optical lattices \cite{jo2012}, see also \cite{Ruostekoski2009}.
Since in experiments it is possible to study bosons in flat bands,
the theoretical interest in bosonic systems with flat bands started
about the same time. 

In a seminal paper, Huber and Altman \cite{Huber2010} studied one-
and a two-dimensional flat band systems with weakly interacting bosons.
It turned out that it is important to understand the detailed structure
of the single-particle eigenstates of the flat band. This is in contrast
to fermions, where the detailed knowledge of the single-particle eigenstates
of the flat band is less important, see e.g. the proofs in \cite{Mielke1999a,Mielke1999,mielke2012},
which use only properties of the projector onto the flat band. In
a flat band it is often possible to construct single-particle eigenstates
that are localised on a finite set of lattice sites. This is also
the case in the systems studied by Huber and Altman \cite{Huber2010}.
They showed that depending on the structure of the lattice and the
structure of the local single-particle eigenstates, the systems may
have different physical properties. Huber and Altman \cite{Huber2010}
investigated the formation of a Bose-Einstein condensate and a charge
density wave. The saw-tooth chain they studied shows the formation
of domain walls and a Luttinger liquid behaviour. Other authors showed
that pairs of bosons are formed in flat band systems \cite{1309.6329v2,Tovmasyan2013,Phillips2014,Pudleiner2015}
and other highly correlated phases \cite{Gremaud2016}.

It is clear that the flat band models must have more than one band,
otherwise the physics of the system would be trivial. Therefore it
is clear that the interesting phenomena observed in flat band systems
occur because other bands are present as well, although they are energetically
often less important. The reason is that the projector onto the flat
band then has a certain structure which leads to effective interactions
of longer range if one projects the interaction part of the Hamiltonian
to the flat band. Huber and Altman \cite{Huber2010} made use of this
projection to construct an effective Hamiltonian with a more complicated
structure which is responsible for the interesting physical properties
they found. The main reason for these interactions is that the projector
onto the flat band falls of algebraically in lattice space, the same
is true for the Wannier states of the flat band. But it is clear that
a projection onto the flat band is in essence a perturbative treatment
of the system that is justified only for weak interactions.

In this paper we are interested in the opposite limit, strong interactions.
We study hard-core bosons on a lattices with flat bands. The questions
we want to answer are whether the phenomena predicted for weakly interacting
systems occur for strong interactions as well and which new phenomena
one can expect. Models with hard-core bosons can be mapped to spin-1/2
systems, since only two states per lattice site are allowed. In fact,
the model we discuss in the present paper is equivalent to such a
spin system in a strong magnetic field. The localised states in the
flat band directly translate to localised magnons and multi-magnon
states which have been investigated intensively, see the review of
Derzhko et al. \cite{Derzkho2007} and the references therein. The
upper critical density appears in those systems as well, see e.g.
\cite{Zhitomirsky2004,Schmidt2006}. For a class of flat band systems,
line graphs of plane graphs, Motruk and Mielke \cite{Motruk2012}
showed rigorously how all multi-particle eigenstates at and below
that critical density can be constructed. In \cite{Schmidt2006},
the linear independence of a subclass of the complete set on some
of those lattices was shown in the context of spin systems. Since
the bosons sit in non-overlapping single particle states, the interaction
strength plays no essential role, all the results apply to hard core
bosons as well. The kagomé lattice and some other lattices mentioned
above fall into this class. The question therefore is, what happens
at densities above the critical density. The pair formation mentioned
above occurs above the critical density and was found in some one-dimensional
systems with stronger interactions \cite{1309.6329v2,Tovmasyan2013,Phillips2014,Pudleiner2015}. 

The paper is organised as follows. The next section contains some
definitions. We fix the notation and introduce some notions from graph
theory needed. We also explain details about the critical density,
following \cite{Motruk2012}. In section \ref{sec:One-dimensional-systems}
we treat a special class of one-dimensional systems, with the chequerboard
chain and the kagomé chain as examples. Both are line graphs. Both
have a local reflection symmetry which allows us to obtain exact results.
We show that pairs are formed and that the pairs are localised. In
section \ref{sec:Two-dimensional-systems} the two-dimensional kagomé
lattice and the chequerboard lattice are discussed. Here as well we
find strong indications for pair formation. The results are based
on the numerical diagonalisation of small systems and on a variational
ansatz. Finally we summarise and discuss our results. 

\section{Definitions}

The Hubbard model is usually defined on a lattice or more general
on a set of vertices $V$. In this paper we consider a bosonic Hubbard
model with bosons which have no internal degree of freedom. The Hamiltonian
can be written in the form
\begin{equation}
H=\sum_{x,y\in V}t_{xy}b_{x}^{\dagger}b_{y}+\sum_{x\in V}U_{x}b_{x}^{\dagger}b_{x}^{\dagger}b_{x}b_{x}\label{eq:H}
\end{equation}
where $T=(t_{xy})_{x,y\in V}$ is the hopping matrix which describes
the hopping of the bosons in a tight binding picture and $U_{x}>0$
are the local, repulsive interactions. We use the usual notation with
creation operators $b_{x}^{\dagger}$ and annihilation operators $b_{x}$
with the usual bosonic commutation relations $[b_{x},b_{y}]=[b_{x}^{\dagger},b_{y}^{\dagger}]=0$
and $[b_{x},b_{y}^{\dagger}]=\delta_{xy}$. 

We will treat this Hamiltonian on a special class of lattices, line
graphs of plane graphs, which we define in the next subsection. The
important property of these lattices is that the lowest eigenvalue
of the hopping matrix is highly degenerate. In the case of translationally
invariant lattices, the lowest band is flat. But this is not the only
important property of these systems.

Further, we are interested in the limit of hard-core bosons, i.e.
we assume that the limit $U_{x}\rightarrow\infty$ is taken. This
means that on each vertex we may have at most one particle. We introduce
the projector $P_{\leq1}$ which projects onto the subspace of the
multi particle Hilbert space which fulfils that property. The Hamiltonian
can then be written as 
\begin{equation}
H=P_{\leq1}\sum_{x,y\in V}t_{xy}b_{x}^{\dagger}b_{y}P_{\leq1}\label{eq:Hprojected}
\end{equation}

\subsection{Line graphs of plane graphs}

We consider the Hubbard Hamiltonian on a graph $G=(V(G),E(G))$. $V(G)$
is the set of vertices, $E(G)$ the set of edges of $G$. An edge
connects exactly two vertices. We consider only undirected graphs,
therefore an edge $e\in E(G)$ is a subset of $V(G)$ with exactly
two elements. The hopping matrix elements in \eqref{eq:H} are $t_{xy}=1$
if $\{x,y\}\in E(G)$, 0 otherwise.

In a graphical representation, the vertices are drawn as dots, connected
with lines, the edges. A graph that can be drawn in a plane without
crossings of edges is called a plane graph. For details we refer to
\cite{Bolobas79}. The edges of the plane graph divide the plane into
many finite and one infinite part, which are called faces. We denote
the set of bounded faces by $F(G)$. Single bounded faces are denoted
by $\lambda\in F(G)$. We denote by $\lambda$ the face and also the
set of edges of the boundary of $\lambda$. The boundary of a face
is an elementary cycle of $G$.

In \cite{Motruk2012} it was shown rigorously that for a certain class
of lattices, line graphs of bipartite plane graphs, there exists a
critical particle number $N_{c}$. For a system with $N\leq N_{c}$
hard-core bosons, it is possible to describe all ground states completely.
In this section we review some of the definitions and results of \cite{Motruk2012}
which we need in the following.

Let $G=(V(G),E(G))$ be a connected bipartite plane graph. Bipartite
means that $V(G)=V_{1}\cup V_{2}$ with $V_{1}\cap V_{2}=\emptyset$
and $|e\cap V_{i}|=1$ for all $e\in E(G)$ and $i=1,2$. If two vertices
are connected, they are not on the same subset. 

According to Eulers theorem, a plane graph has $|F(G)|=|E(G)|-|V(G)|+1$
bounded faces. 

To a graph $G$ one associates various matrices. The most important
is the adjacency matrix $A(G)=(a_{xy})_{x,y\in V(G)}$ where $a_{xy}=1$
if $\{x,y\}\in E(G)$, $a_{xy}=0$ otherwise. A second important matrix
is the vertex-edge incidence matrix $B(G)=(b_{xe})_{x\in V(G),e\in E(G)}$
where $b_{xe}=1$ if $x\in e$, $b_{xe}=0$ otherwise. 

The line graph $L(G)=(V(L(G),E(L(G)))$ of $G$ is constructed as
follows: $V(L(G))=E(G)$, $E(L(G))=\{\{e,e'\}:\,e,e'\in E(G)\,\mbox{and}\,|e\cap e'|=1\}$.
It is easy to see that $A(L(G))=B(G)^{\dagger}B(G)-2$. As a consequence,
the lowest eigenvalue of $A(L(G))$ is $-2$ and it is $|F(G)|$ fold
degenerate. Let us remark that if $G$ is a tree, $|F(G)|=0$ and
there is no state with energy -2. As shown in \cite{Mielke1992a},
a basis of the eigenstates for the eigenvalue $-2$ can be constructed
using the boundaries of the bounded faces $\lambda\in F(G)$ of $G$.
Each bounded face is bounded by an elementary cycle of even length,
because the graph $G$ is bipartite. Each face can be oriented clockwise.
Each edge of $G$ can be oriented to point from one of the two subsets
of $V(G)$ to the other. Now let $v_{\lambda}=(v_{\lambda e})_{e\in E(G)}$
be defined for a face $\lambda$ of $G$ as follows: $v_{\lambda e}=1$
if $e\in\lambda$ and $e$ and $\lambda$ have the same orientation,
$v_{\lambda e}=-1$ if $e\in\lambda$ and $e$ and $\lambda$ have
the opposite orientation, $v_{\lambda e}=0$ otherwise. It is easy
to see that $A(L(G))v_{\lambda}=-2v_{\lambda}$. Since there are $|F(G)|$
bounded faces, which form a set of linear independent $v_{\lambda}$,
these states form a basis of the eigenstates for the eigenvalue $-2$.
For later use we define creation operators $b_{\lambda}^{\dagger}=\frac{1}{\sqrt{|\lambda|}}\sum_{e\in\lambda}v_{\lambda e}b_{e}^{\dagger}$
and the corresponding annihilation operators.

Important examples of line graphs in two dimensions are the kagomé
lattice, the line graph of the honeycomb lattice, and the chequerboard
lattice, the line graph of the square lattice. For further details
and examples we refer to \cite{Mielke1991,Mielke1992a}.

\subsection{Multi-particle states}

Consider now the Hamiltonian \eqref{eq:Hprojected} with the hopping
matrix $A(L(G))$ on the graph $L(G)$ with $N$ hard-core bosons.
Let $C\subset F(G)$ be a subset of non edge-sharing faces. It is
then clear that the state $\prod_{\lambda\in C}b_{\lambda}^{\dagger}|0\rangle$
is a ground state of $H$ with energy $-2|C|$. Clearly, there is
some maximal particle number $N_{c}$ for which that construction
is possible. Further, the states constructed in this way are not complete.
In \cite{Motruk2012} it was shown how all ground states with $N\leq N_{c}$
can be constructed. Further, $N_{c}$ is related to the colouring
of the faces of $G$. Since for $N=N_{c}$ all occupied faces have
no edge in common, they can be coloured with one colour. On the other
hand, if there was a colouring of the faces of $G$ where more than
$N_{c}$ faces are coloured with the same colour, we could choose
this colour and put a particle on each face. But then there would
be a state with energy $-2N$ and since $N\leq N_{c}$ we would have
a contradiction. Therefore, $N_{c}$ is the number of elements in
the largest subset of faces of $G$ that can be coloured with the
same colour. The question we wish to address in this paper is what
happens if one adds a small number of additional particles to this
system. 

Whereas for $N\leq N_{c}$, a mathematically rigorous description
of the ground states is possible, there are no rigorous results for
$N>N_{c}$ so far. To gain some insight here, we use different approaches,
a variational ansatz and numerical diagonalisation of small systems.
For both cases, we need to restrict ourselves to special lattices.
We consider one- and two-dimensional examples derived from the chequerboard
and the kagomé lattice.

\section{One dimensional systems\label{sec:One-dimensional-systems}}

In this section we restrict ourselves to planar bipartite graphs with
the additional property that in a colouring of faces, all elements
of $F(G)$ can be coloured by the same colour. Any two faces have
at most one vertex in common. This class of one-dimensional systems
are chains of even cycles or single edges having a single vertex in
common. Since the graphs in the class are chains of faces, we enumerate
the faces by numbers $1$ to $|F(G)|$ such that the neighbouring
faces of the face $\lambda$ are $\lambda\pm1$.

For each element in this class, we have $N_{c}=|F(G)|$. This simplifies
the proof of completeness of multi-particle states for $N\leq N_{c}$
a lot and makes calculations for $N>N_{c}$ easier. For $N=N_{c}$
there is one single ground state $\prod_{\lambda\in F(G)}b_{\lambda}^{\dagger}|0\rangle$.

The Hamiltonian of hard core bosons on the line graph $L(G)$ can
be written in the form
\begin{equation}
H=H'+H''=\sum_{\lambda\in F(G)}H_{\lambda}+\sum_{\lambda\in F(G)\setminus\{|F(G)|\}}H_{\lambda,\lambda+1}\label{eq:chequerboardChain}
\end{equation}

where $H_{\lambda}$ contains hoppings on the edges of the face $\lambda$
and $H_{\lambda,\lambda+1}$ contains the hoppings between the neighbouring
faces $\lambda$ and $\lambda+1$. To fix our notation, we write
\begin{equation}
H_{\lambda}=P_{\leq1}\sum_{i\in\lambda}b_{\lambda,i}^{\dagger}b_{\lambda,i+1}P_{\leq1}+\mathrm{h.c.}\label{eq:H_lambda}
\end{equation}

where the edges of $\lambda$ are numbered in a cyclic manner by an
integer $i$ modulo $|\lambda|$. As introduced above, $P_{\leq1}$
is the projector onto the subspace with at most one boson on each
vertex of $L(G)$, reflecting the hard core interaction.

Two neighboured faces $\lambda$ and $\lambda+1$ have at most one
vertex in common. If they have no vertex in common, there is exactly
one vertex in $\lambda$ and one vertex in $\lambda+1$ which are
connected by an edge or a chain of edges. Let $i,i+1$ be the two
edges of $\lambda$ connected to the vertex that connects $\lambda$
to $\lambda+1$. By construction, since $L(G)$ is a line graph, $H_{\lambda,\lambda+1}$
is symmetric against an exchange of the two edges $i$ and $i+1$.
An operator that exchanges the two edges is a reflection operator
of the face $\lambda$. We denote this operator by $S_{\lambda}$.
We restrict ourselves to graphs $G$ where $H$ commutes with all
local reflections $S_{\lambda}$, $\lambda\in F(G)$. 

Since all $S_{\lambda}$ commute among themselves and with $H$, all
eigenstates of $H$ are eigenstates of $S_{\lambda}$ as well. Each
eigenstate of $H$ can therefore be characterised by a signature $(s_{\lambda})_{\lambda\in F}$,
where $s_{\lambda}$ takes the values $\pm1$. For the above defined
operator $b_{\lambda}^{\dagger}=\frac{1}{\sqrt{|\lambda|}}\sum_{e\in\lambda}v_{\lambda e}b_{e}^{\dagger}$
, the application of the reflection operator yields $S_{\lambda}b_{\lambda}^{\dagger}S_{\lambda}=-b_{\lambda}^{\dagger}$.
Therefore, the unique ground states for $N=N_{c}$, $\prod_{\lambda\in F(G)}b_{\lambda}^{\dagger}|0\rangle$,
has eigenvalues $-1$ for all $S_{\lambda}$. 

\begin{figure}
\centering{}\includegraphics[width=0.7\columnwidth]{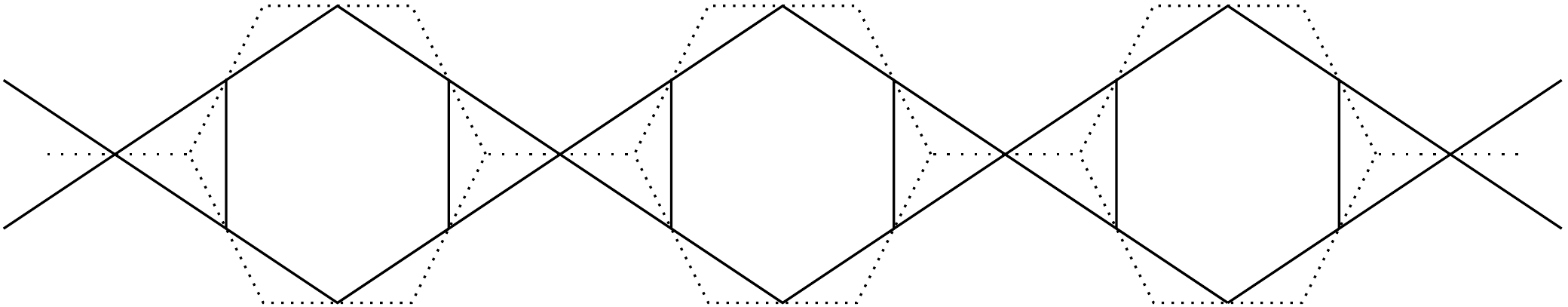}\caption{\label{fig:kagomechain}A part of the kagomé chain, the line graph
of a chain of hexagons connected by additional edges.}
\end{figure}

Let us now add a particle to the system, so that $N=N_{c}+1$. Let
us first consider only $H'$. We obtain ground states for $H'$ by
putting the additional particle on one of the faces $\lambda$. The
resulting ground state has thus two particles on one face $\lambda$
and $s_{\lambda}=1$ for that face, $s_{\lambda}=-1$ for all other
faces. If the elementary cycles $\lambda$ have different length,
it is energetically favourable to put the two particles on the longest
cycle. Let $\ell=\max_{\lambda\in F(G)}|\lambda|$ be the length of
the longest elementary cycle. If there are several cycles of the same
length, the ground state of $\sum_{\lambda}H_{\lambda}$ will be degenerate.
The most interesting case is clearly the one where all cycles have
the same length, in that case the ground state $\sum_{\lambda}H_{\lambda}$
with $N=N_{c}$ particles will be $N_{c}$-fold degenerate.The question
is now, what happens if we add the hopping between the faces, $\sum_{\lambda}H_{\lambda,\lambda+1}$.
In that case, the pair of particles sitting on one of the faces can
gain some energy by spreading to the neighboured faces, the ground
state energy gets lower. But, since all $H_{\lambda,\lambda+1}$ obey
the same symmetry, we may expect to get still a ground state with
a signature, where all $s_{\lambda}$ are $-1$ except one, which
has $+1$. This idea has an immediate problem: Suppose that there
are many edges between two neighboured faces $\lambda$ and $\lambda+1$.
Then, spreading the additional particle on these edges, the particle
can gain an energy close to $-2$, which is the energy on an infinitely
long chain. Therefore we restrict ourselves to chains with at most
a single edge between two faces. To proceed, we treat two examples,
the chequerboard chain and a kagomé chain. The chequerboad chain is
the line graph of a chain of vertex sharing squares. The kagomé chain
is the line graph of a chain of disjoint hexagons connected by additional
edges, one between each pair of neighboured hexagons see Fig. \ref{fig:kagomechain},
which also shows its construction as a line graph. 

Let us mention that other kagomé chains have been proposed and treated
before, esp. in the context of frustrated spin systems, see \cite{Azaria1998,Waldtmann2000,Schulenburg2002}.
None of those falls into the class of models discussed here. These
one-dimensional lattices are line graphs; the kagomé like chain of
\cite{Azaria1998} is the line graph of a chain of edge sharing squares,
the one in \cite{Waldtmann2000} a chain of edge sharing hexagons,
both are discussed in \cite{Schulenburg2002}. For these models, $N_{c}=|F(G)|/2$,
not $N_{c}=|F(G)|$ as in the class of models we are treating here.

\subsection{\label{subsec:The-chequerboard-chain}The chequerboard chain}

\begin{figure}
\centering{}\includegraphics[width=0.5\columnwidth]{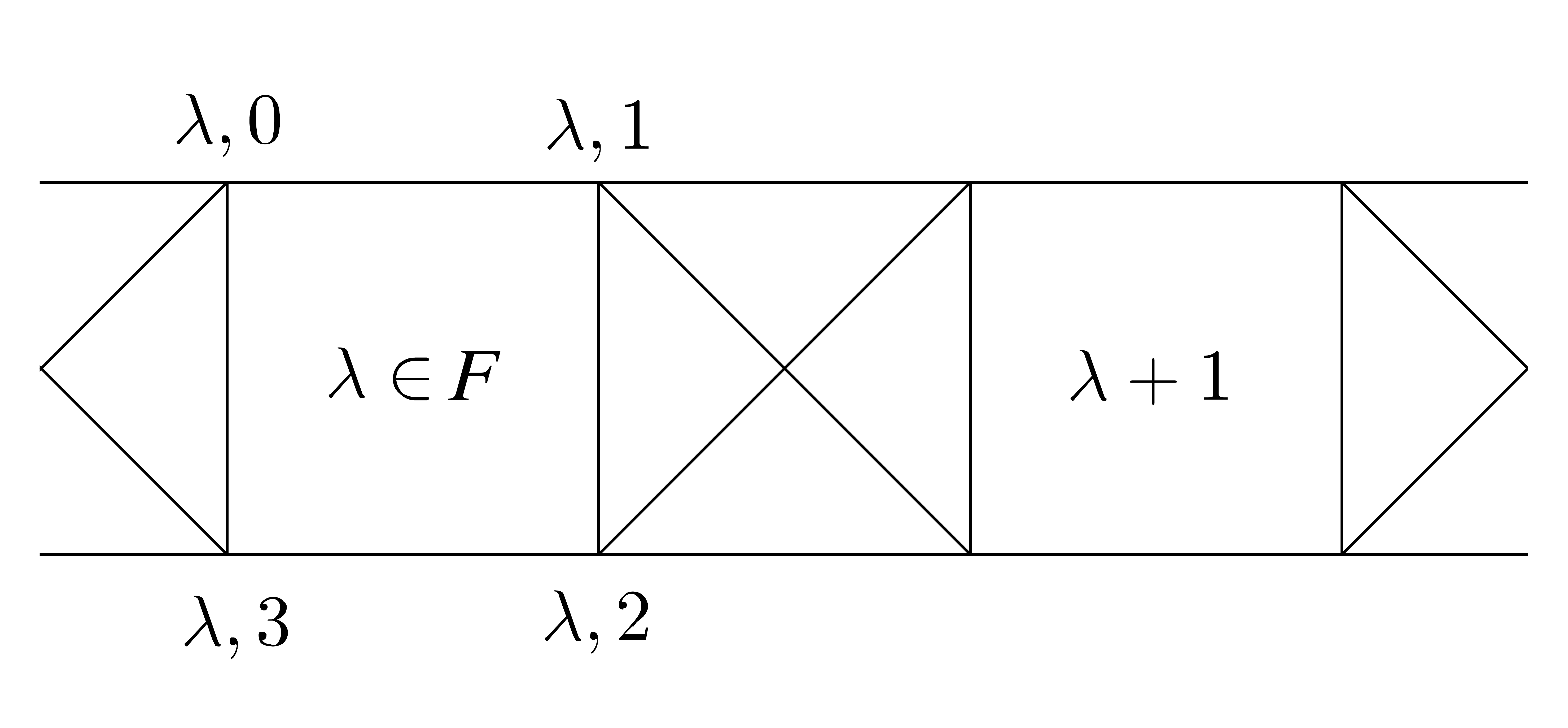}\caption{\label{fig:2_lin_Bezeichnungen}A part of the chequerboard chain,
the line graph of a chain of corner sharing squares.}
\end{figure}

Let $G$ be a chain of squares which have one vertex in common. The
line graph of $G$ is the chequerboard chain. A part of the chequerboard
chain $L(G)$ is depicted in Fig. \ref{fig:2_lin_Bezeichnungen}.
Each face contains four edges of $G$ numbered as depicted in \ref{fig:2_lin_Bezeichnungen}.
With this notation the Hamiltonian can be written as in \eqref{eq:chequerboardChain}
with
\begin{equation}
H_{\lambda,\lambda+1}=P_{\leq1}(b_{\lambda,1}^{\dagger}+b_{\lambda,2}^{\dagger})(b_{\lambda+1,0}+b_{\lambda+1,3})P_{\leq1}+\mathrm{h.c.}\label{eq:H_lambda,lambda+1}
\end{equation}

The operator that performs the local reflection has the explicit form
\begin{equation}
S_{\lambda}=\exp(\frac{i\pi}{2}((b_{\lambda,1}^{\dagger}-b_{\lambda,2}^{\dagger})(b_{\lambda,1}-b_{\lambda,2})+(b_{\lambda,0}^{\dagger}-b_{\lambda,3}^{\dagger})(b_{\lambda,0}-b_{\lambda,3}))).\label{eq:localReflection}
\end{equation}
To verify that, notice first that since $b_{\lambda,1}^{\dagger}+b_{\lambda,2}^{\dagger}$
commutes with $b_{\lambda,1}-b_{\lambda,2}$, it commutes also with
$S_{\lambda}$. Further, $S_{\lambda}^{\dagger}(b_{\lambda,1}^{\dagger}-b_{\lambda,2}^{\dagger})S_{\lambda}=\exp(i\pi)(b_{\lambda,1}^{\dagger}-b_{\lambda,2}^{\dagger})=-(b_{\lambda,1}^{\dagger}-b_{\lambda,2}^{\dagger}).$
Therefore, $S_{\lambda}^{\dagger}b_{\lambda,1}^{\dagger}S_{\lambda}=b_{\lambda,2}^{\dagger}$,
and similarly for $b_{\lambda,0}^{\dagger}$ and $b_{\lambda,3}^{\dagger}$. 

We start with open boundary conditions and let $N=N_{c}+1$, i.e.
we put one additional particle in the system. The results of a numerical
diagonalisation of small systems is shown in table \ref{tab:lin_untersch_lang}.
The second column contains $\Delta E=E+2N_{c}$. For longer chains
the lowest eigenvalue stays at -1.0165. A trivial upper bound for
$\Delta E$ is $\Delta E<-2\sqrt{2}+2=-0.8284$. Since $-2\sqrt{2}$
is the ground state energy of two hard-core bosons on a square, the
ground states of $H'$ have exactly $\Delta E=-2\sqrt{2}+2$. Since
the ground state of a doubly occupied square has $s_{\lambda}=1$,
the ground state of $H'$ is exactly $|F(G)|$-fold degenerate. Adding
$H''$ to $H'$ lowers the energy of the eigenstates. Note that for
the lattices in table \ref{tab:lin_untersch_lang} there are exactly
$N_{c}$ states with $\Delta E<-2\sqrt{2}+2$. 

Looking at the states shown in table \ref{tab:lin_untersch_lang}
shows that they have a strong overlap with states where one face is
occupied by two hard-core bosons and all the other faces are occupied
by one particle. Adding $H''$ to $H'$ has the effect that the additional
particle spreads slightly on the other faces, thereby lowering the
energy of the pair. But it does not change the symmetry properties
of the ground state, the signature $(s_{\lambda})_{\lambda\in F(G)}$.
The reason is that the ground states of $H'$ have a non-zero overlapp
with the low energy states of $H$, since $H''$ respects the symmetry
and higher energy states of $H'$ are not lowered enough to become
new ground states of $H$. The signatures of the states are $-1$
on all faces except one, which has $+1$. This face is the doubly
occupied face. Only at the boundary the lowering is somewhat less,
therefore we obtain two states with an energy -0.920 for the lattices
in table \ref{tab:lin_untersch_lang}. This is due to the fact, that
the two-particle ground state on a single square has signature $+1$,
so there must be a state with an energy lower than the trivial one
in each of these $N_{c}$ subspaces and by table \ref{tab:lin_untersch_lang},
only $N_{c}$ such states exist.

\begin{table}
\centering{}%
\begin{tabular}{|c|l|}
\hline 
lattice  & %
\begin{minipage}[b]{0.15\textwidth}%
all eigenvalues ($\Delta E$) 

below$-2\sqrt{2}+2$%
\end{minipage}\tabularnewline
\hline 
\hline 
\includegraphics[height=1cm]{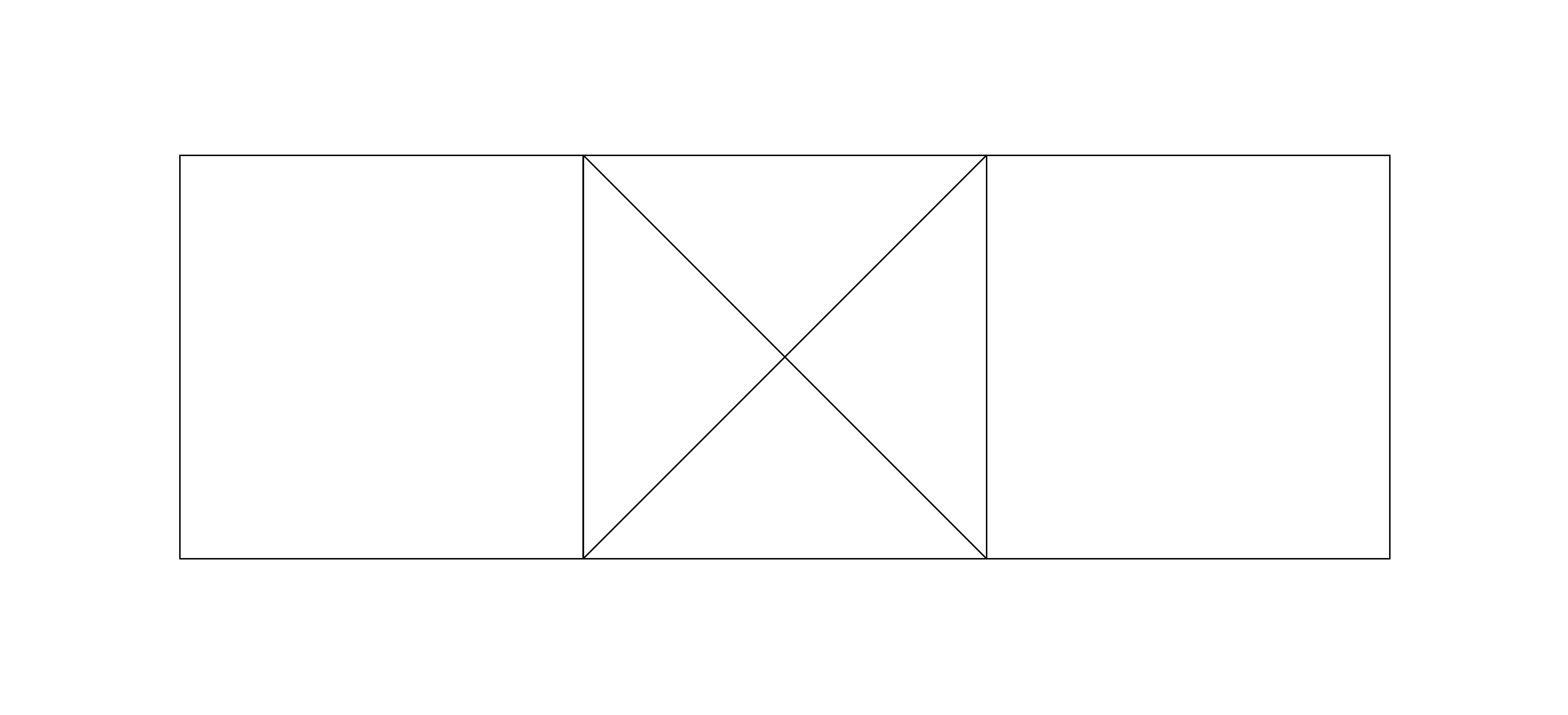} & %
\begin{minipage}[b]{0.19\textwidth}%
-0.920; -0.920

~%
\end{minipage}\tabularnewline
\hline 
\includegraphics[height=1cm]{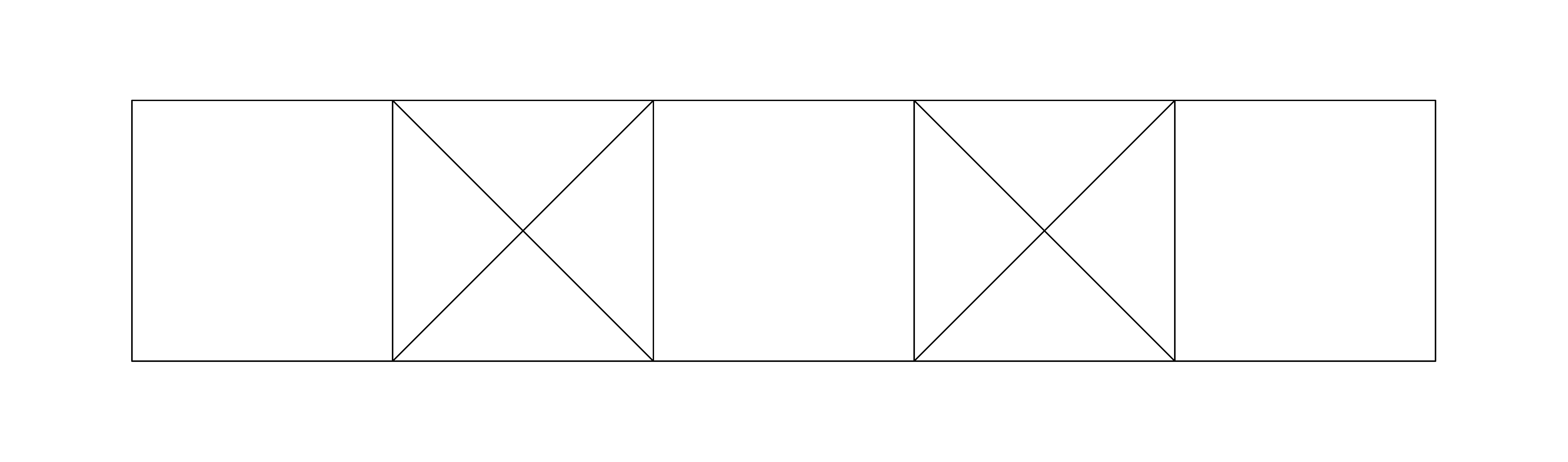} & %
\begin{minipage}[b]{0.15\textwidth}%
-1.0162\\
-0.920; -0.920%
\end{minipage}\tabularnewline
\hline 
\includegraphics[height=1cm]{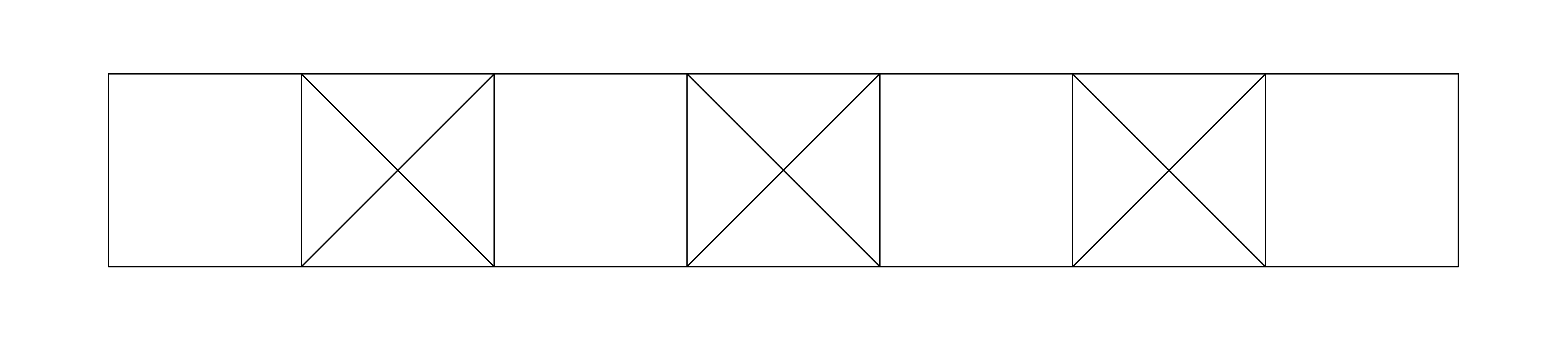} & %
\begin{minipage}[b]{0.19\textwidth}%
-1.0163; -1.0163\\
-0.920; -0.920%
\end{minipage}\tabularnewline
\hline 
\includegraphics[height=1cm]{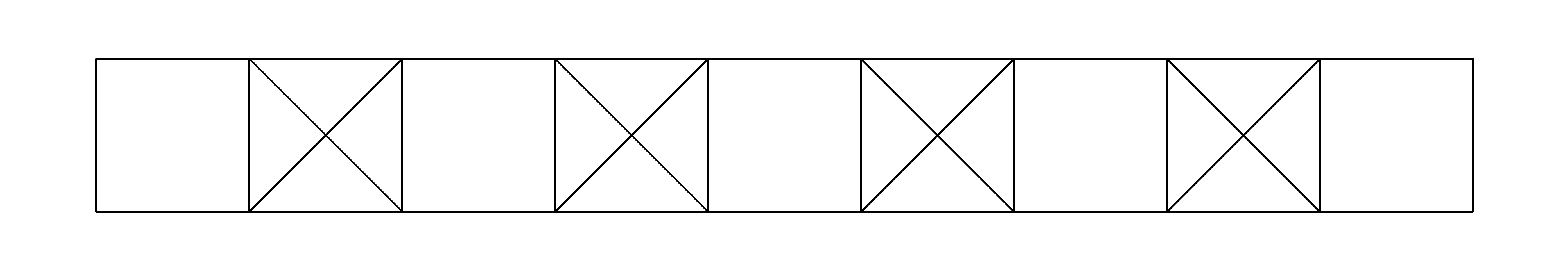} & %
\begin{minipage}[c][1\totalheight][b]{0.19\textwidth}%
-1.0165\\
-1.0164; -1.0164\\
-0.920; -0.920%
\end{minipage}\tabularnewline
\hline 
\includegraphics[height=1cm]{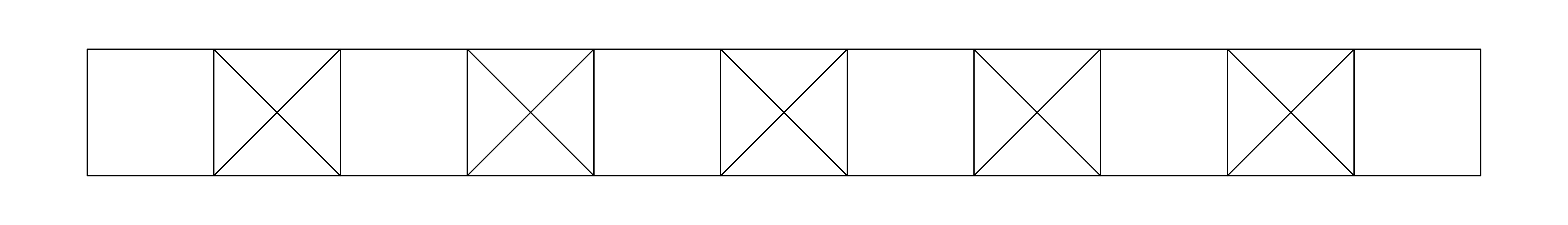} & %
\begin{minipage}[c][1\totalheight][b]{0.19\textwidth}%
-1.0165; -1.0165\\
-1.0164; -1.0164\\
-0.920; -0.920%
\end{minipage}\tabularnewline
\hline 
\end{tabular}\caption{\label{tab:lin_untersch_lang}Eigenvalues of finite open parts of
the chequerboard chain with $N_{c}+1$ bosons. The eigenvalues are
shown relative to the ground state energy of the system with $N_{c}$
bosons, $\Delta E=E+2N_{c}$.}
\end{table}

From the ground state one can calculate the occupation numbers on
the faces. The number of particles on the face $\lambda$ is $n_{\lambda}=\sum_{i\in\lambda}n_{\lambda,i}=\sum_{i\in\lambda}b_{\lambda,i}^{\dagger}b_{\lambda,i}$.
For the ground state of the lattice with five faces we obtain $\langle n_{2}\rangle=1.91$,
$\langle n_{1}\rangle=\langle n_{3}\rangle=1.04$, $\langle n_{0}\rangle=\langle n_{4}\rangle=1.000004$.
For longer chains, the density on more distant squares is 1 within
the numerical accuracy. This shows that the additional boson forms
a pair on one face and that the pair is strongly localised. It is
spread a bit on the nearest neighboured faces and to a very low portion
to the next neighboured faces.

Let us now assume that we have periodic boundary conditions and an
arbitrary $N$. Let us assume that we have an eigenstate of $H$ with
a signature $(s_{\lambda})_{\lambda\in F(G)}$. Then, because of the
translational symmetry of the Hamiltonian, the state that is translated
by 1 face is as well an eigenstate, it has the signature $(s_{\lambda}')$
with $s_{\lambda}'=s_{\lambda+1}$. As a consequence, we get degeneracies
in the spectrum of $H$. The degeneracies are multiples of divisors
of $|F(G)|$. 

For a chain with periodic boundary conditions and $N=N_{c}+1$ we
obtain degenerate ground states at energy -1.0165. As for the open
chain, the signatures of the states are $-1$ on all faces except
one, the doubly occupied face, which has $+1$. Due to the translational
symmetry, there are exactly $|F(G)|$ degenerate states, which are
found in the numerical diagonalisation of systems with small $|F(G)|$.
The reasoning is the same as above, the ground states of $H$ have
a finite overlap with the ground states of $H'$ and therefore must
have the same symmetry properties, which directly yields the $|F(G)|$-fold
degeneracy.

This analysis is not rigorous in a mathematical sense because it relies
on numerical diagonalisations of small systems. But since the symmetry
is exact, the localisation of the eigenstates is very strong, and
the eigenvalues do not change for longer chains, we believe to have
a very strong argument that the analysis holds true for arbitrary
long chains. The physical picture for additional hard-core bosons
added to the chequerboard chain is that we obtain an effective flat
band for these particles, that for $n\ll|F(G)|$ additional particles
the ground state energy is $E=-2|F(G)|-1.0165n$ and that the additional
particles form localised pairs with the particles already present.
We checked that numerically for $n=2$. The two pairs try to separate.
On sufficiently long chains, the ground state energy is indeed $E=-2|F(G)|-2\times1.0165$.
For shorter chains, the pairs start to overlap and the energy rises.
This is physically intuitive: due to the spread of the pair state
to neighboured faces, the pairs will try to keep some distance from
each other, having an effective repulsive short-range interaction. 

The derivation of this physical picture relies on the local reflection
symmetry combined with the translation symmetry of the chequerboard
chain. The question is whether a similar picture is true for other
one-dimensional lattices without a local reflection symmetry or even
for two dimensional systems.

\subsection{Kagomé chain}

The kagomé chains we treat here is the line graph $L(G)$ of a graph
$G$ that is a chain of hexagons connected by a single edge between
neighboured hexagons, see Fig. \ref{fig:kagomechain}. This kagomé
chain contains therefore interstitial sites, which are important for
weakly interacting bosons on the two-dimensional kagomé lattice \cite{Huber2010}.
For this chain, we can proceed as for the chequerboard chain, and
the results are similar. We therefore do not repeat each argument
above, but only state the results. For chains with periodic boundary
conditions, we obtain $|F(G)|$-fold degenerate ground states with
an energy $\Delta E=-1.6046$. The states are localised. The occupation
numbers are 1.80 for the central hexagon, 0.086 for the neighboured
interstitial sites, and 1.014 for the neighboured hexagons. Within
the numerical accuracy the density on more distant hexagons is 1.
For longer open chains, the ground states also have the same energy
starting with a chain of three hexagons. This and the fact that even
for three hexagons $\Delta E$ has the energy of the chain with periodic
boundary conditions indicate that the localisation of the pair is
even stronger than for the chequerboard chain. The physical reason
may be that it is energetically easier to put two hard core bosons
on a hexagon than on a square. For the open chains, we find two states
with a somewhat higher energy $\Delta E=-1.518$ corresponding to
states localised on the two outer hexagons. 

As for the chequerboard chain, we get an effective flat band for the
additional particle. Due to the strong localisation, adding $n\ll|F(G)|$
additional particles the ground state energy is $E=-2|F(G)|-1.6046n$
and that the additional particles form localised pairs with the particles
already present. 

As for the chequerboard chain, the results rely on numerical diagonalisations
of small systems. But for the same reasons as above, we expect them
to be true for arbitrary long chains. In both cases, we get an entirely
flat band for the additional particle in a chain with periodic boundary
conditions. The additional particle forms a localised pair. For open
boundary conditions the band is almost flat, there are edge states
with a somewhat higher energy.

\section{Two-dimensional systems\label{sec:Two-dimensional-systems}}

In this section, we treat two-dimensional systems of which the one-dimensional
systems of the previous section are sublattices: The kagomé lattice
and the chequerboard lattice. These two-dimensional systems have no
local reflection symmetry. We base our analysis on variational states
and numerical diagonalisations of small systems.

We start our analysis with the kagomé lattice because it is itself
a plane graph and because it has interstitial sites. Interstitial
sites give some additional space for additional particles above the
critical density. Therefore, some other states, which do not consist
of pairs, may be more favourable. Further, the question of pair formation
on the chequerboard lattice has been discussed already in \cite{Pudleiner2015}
and therefore a lengthy treatment is not necessary here. 

\subsection{The kagomé lattice}

Let $F_{1}$ be one of the three subsets of non edge-sharing faces
of $F(G)$ where $G$ is now the honeycomb lattice and $L(G)$ is
the kagomé lattice. We assume that we have open boundaries and that
$F_{1}$ is the largest of the three subsets of non edge-sharing faces.
Let $F_{1}^{\dagger}=\prod_{\lambda\in F_{1}}b_{\lambda}^{\dagger}$.
The state $|F_{1}\rangle=F_{1}^{\dagger}|0\rangle$ is a ground state
of the system for $|F_{1}|$ particles. If $|F_{1}|=|F(G)|/3$, there
are two more ground state with the same particle number but formed
on one of the other two subsets of non edge-sharing faces. 

The difference between the kagomé lattice and the chequerboard lattice,
see below, is that the former contains interstitial sites, sites which
are not occupied in $|F_{1}\rangle$. Huber and Altman treated the
bosonic Hubbard model on the kagomé lattice for weak interaction and
found that the additional particle gets delocalised, mainly because
of the interstitial sites. 

To investigate the role of the interstitial sites and a possible delocalisation,
we construct a variational state 
\begin{equation}
|X_{\leq1}\rangle=P_{\leq1}\psi^{\dagger}|F_{1}\rangle\label{eq:varX}
\end{equation}
\begin{equation}
\psi^{\dagger}=\sum_{e\in E(G)}\psi_{e}b_{e}^{\dagger}\label{eq:varpsi}
\end{equation}

Denote $|\psi\rangle=\psi^{\dagger}|0\rangle$ the corresponding single
particle state. Then we obtain 
\begin{eqnarray}
\frac{\langle X_{\leq1}|H|X_{\leq1}\rangle}{\langle X_{\leq1}|X_{\leq1}\rangle} & = & \frac{\langle\psi|F_{1}P_{\leq1}HP_{\leq1}F_{1}^{\dagger}|\psi\rangle}{\langle\psi|F_{1}P_{\leq1}F_{1}^{\dagger}|\psi\rangle}\label{eq:var+1}
\end{eqnarray}
Introducing the matrices $M=(m_{ee'})_{e,e'\in E(G)}$ and $C=(c_{ee'})_{e,e'\in E(G)}$
where 
\begin{equation}
m_{ee'}=\langle0|b_{e'}F_{1}P_{\leq1}(H+2|F_{1}|)P_{\leq1}F_{1}^{\dagger}b_{e}^{\dagger}|0\rangle\label{eq:varM}
\end{equation}
and 
\begin{equation}
c_{ee'}=\langle0|b_{e'}F_{1}P_{\leq1}F_{1}^{\dagger}b_{e}^{\dagger}|0\rangle\label{eq:varC}
\end{equation}
we can write the variational energy as 
\begin{eqnarray}
\frac{\langle X_{\leq1}|H|X_{\leq1}\rangle}{\langle X_{\leq1}|X_{\leq1}\rangle} & = & -2|F_{1}|+\frac{\langle\psi|M|\psi\rangle}{\langle\psi|C|\psi\rangle}\label{eq:effectiveVarH}
\end{eqnarray}
The second term in this equation is $\Delta E$, the energy of the
additional particle. $M$ and $C$ can be calculated to be
\begin{align}
M & =H-\frac{1}{3}(HP_{F(G)}+P_{F(G)}H)+\frac{2}{3}P_{F}-2P_{F_{1}}\label{eq:matM}\\
C & =1-\frac{1}{3}P_{F(G)}+P_{F_{1}}\label{eq:matC}
\end{align}
where $(P_{F(G)})_{e,e'}=\delta_{e\in F(G)}\delta_{ee'}$ is the projector
on $F(G)$ and $(P_{F_{1}})_{ee'}=\frac{1}{6}\sum_{\lambda\in F}\nu_{\lambda e}\nu_{\lambda e'}$
is the projector on the one-particle states that $F_{1}$ is build
of. Since eigenvalues for projectors are 0 or 1, $C$ is positive
definite and therefore it is possible to introduce a new scalar product
$\{\phi|\psi\}=\langle\phi|C|\psi\rangle$. Using this notation one
obtains $\Delta E=\frac{\{\psi|C^{-1}M|\psi\}}{\{\psi|\psi\}}$. The
matrix $C^{-1}M$ can be calculated by noting that $C^{-1}=1+\frac{1}{2}P_{F(G)}-\frac{9}{10}P_{F_{1}}$.
Due to the translational invariance of the lattice and the state $|F_{1}\rangle$
we can apply a Fourier transform. For the kagomé lattice, the resulting
matrix which has to be diagonalised is a $9\times9$-matrix. We obtain
$9$ energy bands for the additional particle using the variational
ansatz.

\begin{figure}
\centering{}\includegraphics[width=0.6\columnwidth]{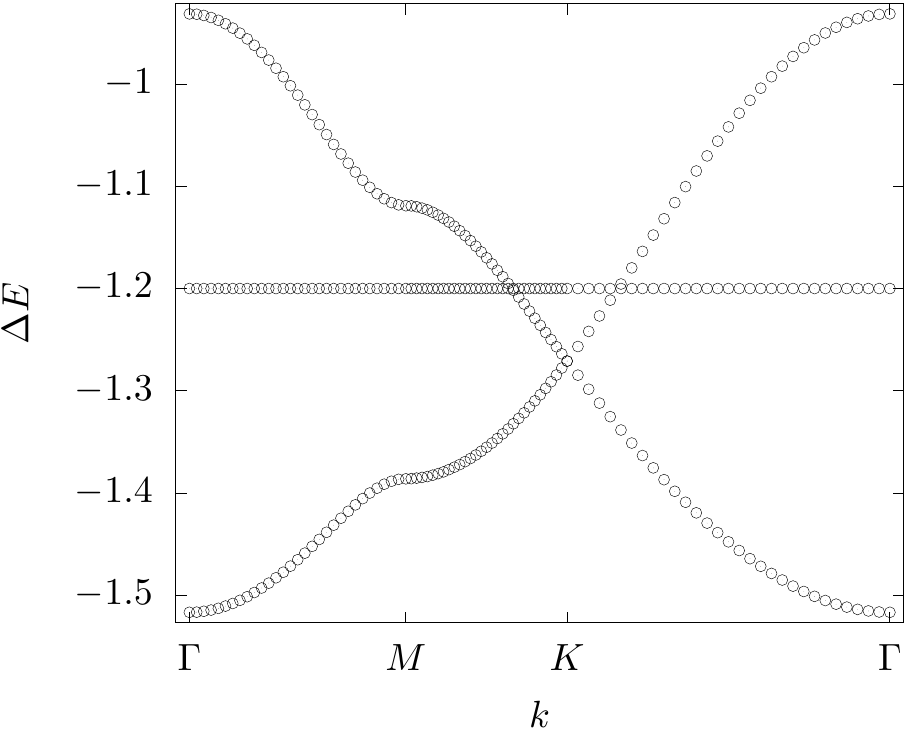}\caption{\label{fig:Lowest-three-energy}Lowest three energy bands of an additional
particle added to the ground state $|F_{1}\rangle$ within the variational
ansatz \eqref{eq:var+1}.}
\end{figure}

In Fig. \ref{fig:Lowest-three-energy} we show the lowest three energy
bands. We obtain two dispersive bands and one flat band, the latter
is not at the bottom of the spectrum. The lowest energy is $\Delta E=-1.52$. 

\begin{figure}
\begin{centering}
\begin{tabular*}{0.8\textwidth}{@{\extracolsep{\fill}}cc}
\includegraphics[width=0.35\textwidth]{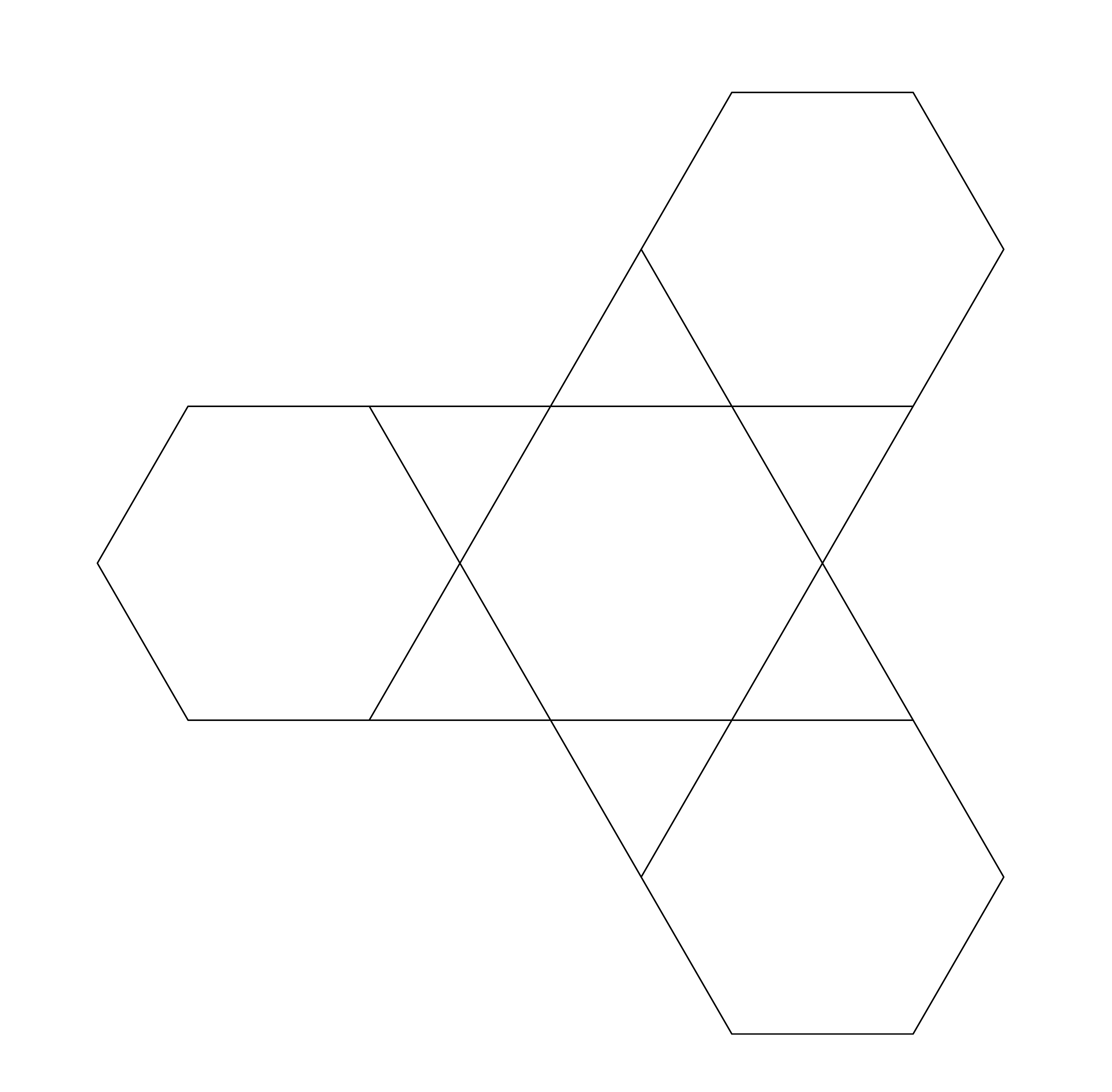} & \includegraphics[width=0.35\textwidth]{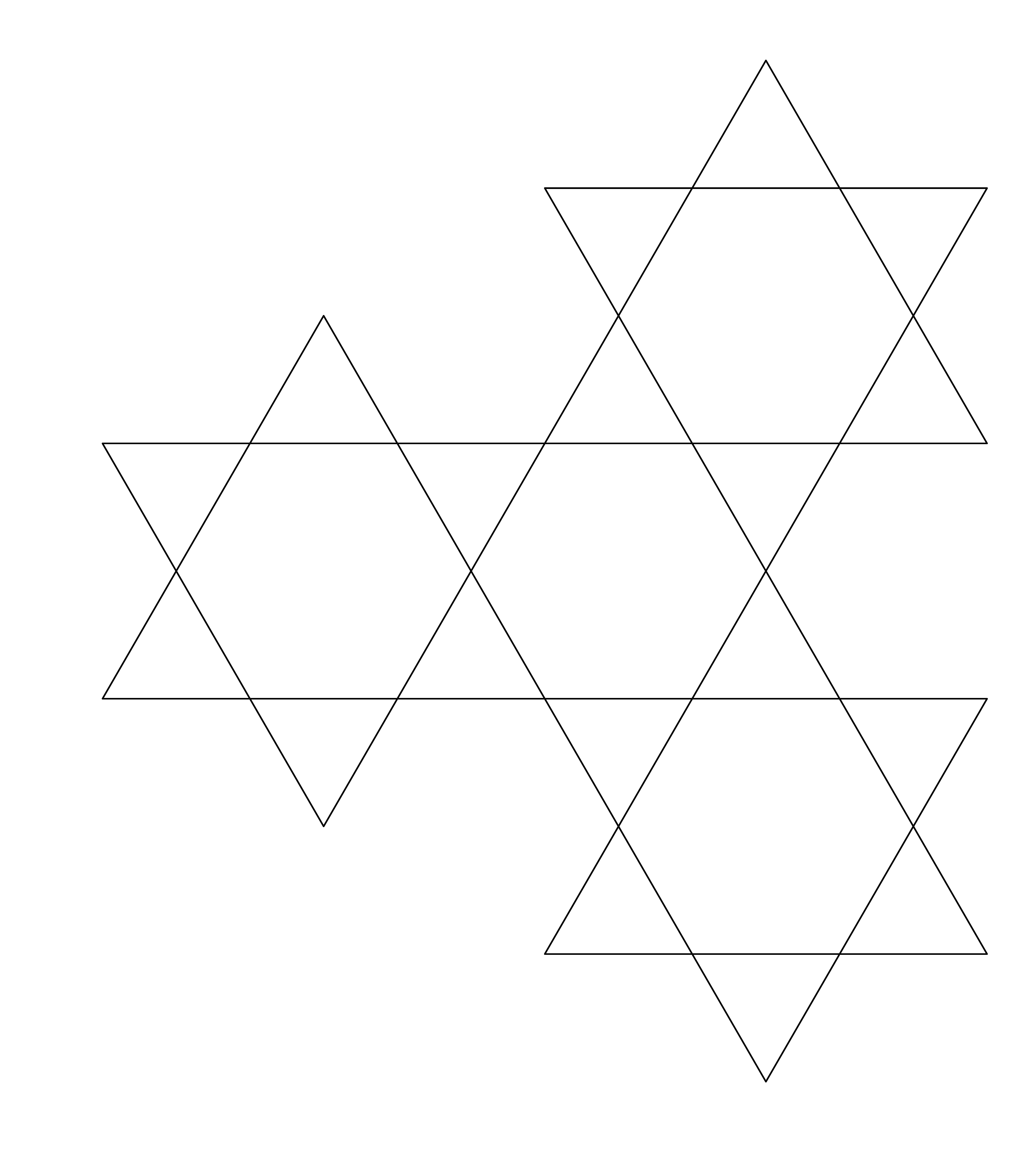}\tabularnewline
\end{tabular*}
\par\end{centering}
\caption{\label{fig:non-decorated-and-decorated}non-decorated and decorated
subunits of the kagomé lattice.}

\end{figure}

In \cite{Pudleiner2015} it was shown that putting an additional particle
onto an already occupied face $\lambda$, one obtains the energy $\mbox{\ensuremath{\Delta}E=}2-2\sqrt{3}=-1.46$.
The variational state constructed above has a lower energy which indicates
that the particle gains some energy due to the delocalisation. On
the other hand, Pudleiner et al. \cite{Pudleiner2015} also considered
the subunit shown in Fig. \ref{fig:non-decorated-and-decorated}.
Adding a particle to that subunit one obtains four states with an
energy below -1.46. Three of them are almost degenerate at energies
$\Delta E=-1.55$ and $\Delta E=-1.56$ corresponding to linear combinations
of the doubly occupied faces, the fourth has the energy $\Delta E=-1.66$.
In this state, the interstitial sites are occupied as well. These
energies are upper bounds to the true energy of the additional particle
on the kagomé lattice. Decorating the subunit with 12 additional sites,
see Fig. \eqref{fig:non-decorated-and-decorated}, one obtains somewhat
lower energies, $\Delta E=-1.69$ and $\Delta E=-1.75$. These are
no upper bounds since the additional sites would touch occupied faces.
The energies are lower, so the additional particle gets further delocalised
onto the additional sites, but the amplitudes on these sites are low,
0.008 and 0.003, showing that the trend to delocalise the particle
further is not high. All these energies are lower than the lowest
energy of the variational ansatz \eqref{eq:var+1}, which shows that
the energy gain due to delocalisation is smaller than the energy gain
one obtains if the particle is in a localised state and arranges itself
in an optimal way with its neighbours. 

\subsection{The chequerboard lattice}

Localised pairs on the chequerboard lattice have been proposed in
\cite{Pudleiner2015}. Diagonalising a subunit of the chequerboard
lattice with five squares and five bosons (one more than $|F(G)|$)
they found four almost degenerate states, the lowest having $E=-2|F(G)|-1.081$,
the additional particle has an energy $-1.081$. This energy is a
variational upper bound of the true ground state energy in a system
with large $|F(G)|$. The four states have a large overlap with the
states where a pair of bosons is localised on one square. The energy
is lowered and the degeneracy is lifted, although only weakly, because
the pair spreads a bit, as for the chain. Note that the energy is
significantly lower than for the chain with five squares, indicating
that the delocalisation is larger here. 

The variational calculation is possible on the kagomé lattice, because
there the matrix $M$ has a relatively simple form. The reason is
that a particle hopping from a site on a face $\lambda$ to some site
not in $\lambda$ hops to an interstitial site. A single hopping never
transfers the particle to a neighboured face. On the chequerboard
lattice this is not true. Therefore, the same variational calculation
on the chequerboard lattice yields an additional term in \eqref{eq:matM}.
The calculation is still possible but does not yield additional insight. 

\section{Summary and Outlook}

The aim of this paper is to investigate hard-core bosons in a flat
band system. We first looked at two one-dimensional systems, a chequerboard
chain and a kagomé chain. The chequerboard chain is the line graph
of a chain of corner sharing squares. The kagomé chain is a line graph
of a chain of hexagons connected by additional edges. As line graphs,
both have a lowest flat band. Further, both have a local reflection
symmetry. Applying the reflection of a single square or hexagon, an
eigenstate of the Hamiltonian is either symmetric or antisymmetric.
At the critical density, each square or hexagon is occupied by a single
particle and the ground state is antisymmetric with respect to all
local reflections. Adding a single particle to the system, the ground
state of the chain becomes symmetric for one local reflection and
remains antisymmetric for all others. In the case of periodic boundary
conditions, we obtain a degeneracy given by the number of squares
or hexagons. For open boundary conditions, most of the states are
almost degenerate except for those where the state is symmetric with
respect to reflection of the first or last square or hexagon. At the
boundary, the additional particle has a slightly higher energy.

In the ground states with one additional particle, this particle forms
a pair with another particle. The lowest state with one particle on
the square or hexagon is antisymmetric, with two particles it is symmetric.
This explains the symmetry of the states and the degeneracy. Looking
closer at the states we find that the pair which is formed is localised.
Effectively, we obtain a flat band for the additional particle. The
localisation is not perfect in the sense that the pair is strictly
localised on a single square or hexagon, but the expectation value
of occupation number falls of very rapidly. For the chains, already
at a distance of three faces, the excess density is 0 within the numerical
accuracy. This means that even at a finite but low density above the
critical density we expect the ground state to be degenerate and to
be formed of well separated localised pairs and single particles in
between. The numerical results for two additional particle on finite
chains support that view.

In two dimensions, the chequerboard and kagomé lattice are line graphs
as well with a lowest flat band. But for the two-dimensional systems,
there is no exact local symmetry. Therefore, it is not possible to
obtain exact results as for the one-dimensional analogues. Here, we
use two variational ansatzes, one with possibly extended states and
one with localised states on small compounds. The best variational
states for the two systems again show localised pairs, although the
localisation is less perfect than in the one-dimensional case. But
we have a strong indication that here as well the additional particle
forms a localised pair and that one obtains an effectively flat band
for the additional particles at low densities.

Since pair formation was observed in other one- and two-dimensional
bosonic flat band systems as well \cite{1309.6329v2,Tovmasyan2013,Phillips2014,Pudleiner2015},
this seems to be a universal feature of these systems. But whether
or not this is really the case remains open. Whereas for fermions
in a flat band a full characterisation of the ground states is possible
at and below half filling of the flat band, this is not the case for
bosons. For fermions, a detailed knowledge of the single-particle
eigenstates is not necessary, often some properties of the projector
are sufficient to obtain the desired result. In contrast, for bosons,
the knowledge of the single particle states and, in the case of the
two one-dimensional systems discussed here, their symmetry is important
to understand the underlying physics of the system. This makes the
treatment of interacting bosons in a flat band much more challenging. 

The kagomé lattice has been realised as an optical lattice \cite{Ruostekoski2009,jo2012}
and experiments are possible which may allow to investigate pair formation.
For the chequerboard lattice an experimental realisation may be difficult
or even impossible. But the flat band states of the chequerboard lattice
are similar to those in the Lieb lattice \cite{Mielke1991}, there
is a one-to-one mapping of the single particle states in the flat
bands of these two lattices. Adding additional on site energies to
the Lieb lattice or next nearest neighbour hoppings, it is even possible
to shift the flat band to the bottom of the spectrum. The Lieb lattice
has been realised as an optical lattice as well \cite{Taie2015}.
Therefore it may be possible to investigate pair formation in the
Lieb lattice. To our knowledge, hard core bosons in a Lieb lattice
have not been studied so far.

Author contribution statement: Both authors contributed equally to
the paper.


\begin{thebibliography}{10}
\bibitem{Hubbard63} J~Hubbard. \newblock {Proc. Roy. Soz. A}
\textbf{276}, 238 (1963).

\bibitem{Kanamori63} J~Kanamori. \newblock {Prog. Theor. Phys.}
\textbf{30}, 275 (1963).

\bibitem{Gutzwiller1963} M.~C. Gutzwiller. \newblock {Phys. Rev.
Lett.} \textbf{10}, 159 (1963).

\bibitem{Fisher1989} M.P.A. Fisher, P.B. Weichman, G.~Grinstein,
D.S. Fisher. \newblock {Phys. Rev. B} \textbf{40}, 546 (1989).

\bibitem{Lieb89} E~H Lieb. \newblock {Phys. Rev. Lett.}\textbf{62},
1201 (1989).

\bibitem{Mielke1991} A~Mielke. \newblock {J. Phys. A: Math. Gen.}
\textbf{24}, 3311 (1991).

\bibitem{Tasaki92} H~Tasaki. \newblock {Phys. Rev. Lett.} \textbf{69},
1608 (1992).

\bibitem{Mielke1993} A. Mielke, H. Tasaki. \newblock {Commun. Math.
Phys.} \textbf{158}, 341 (1993).

\bibitem{Tasaki97b} H~Tasaki. \newblock {cond-mat/9712219} (1997).

\bibitem{Mielke1999} A~Mielke. \newblock {J. Phys. A, Math. Gen.},
\textbf{32}, 8411 (1999).

\bibitem{Schulenburg2002}J. Schulenburg, A. Honecker, J. Schnack,
J. Richter, and H.-J. Schmidt. Phys. Rev. Lett. 88, 167207 (2002).

\bibitem{Derzkho2007}O. Derzhko, J. Richter, A. Honecker, and H.-J-
Schmidt. \newblock {Low Temp. Phys.} \textbf{33}, 745 (2007).

\bibitem{jo2012} G.-B. Jo, J~Guzman, C~K Thomas, P~Hosur, A~Vishwanath,
D~M Stamper-Kurn. \newblock {Phys. Rev. Lett.} \textbf{108}, 045305
(2012).

\bibitem{Ruostekoski2009} J~Ruostekoski. \newblock {Phys. Rev.
Lett.} \textbf{103}, 080406 (2009).

\bibitem{Huber2010} S~D Huber, E~Altman. \newblock {Phys. Rev.
B} \textbf{82}, 184502 (2010).

\bibitem{Mielke1999a} A. Mielke. \newblock {Physical Review Letters}
\textbf{82}, 4312 (1999).

\bibitem{mielke2012} A. Mielke. \newblock {Eur. Phys. J. B} \textbf{85},
1 (2012).

\bibitem{1309.6329v2} S. Takayoshi, H. Katsura, N. Watanabe, H. Aoki.
\newblock {Phys. Rev. A} \textbf{88}, 063613 (2013).

\bibitem{Tovmasyan2013} M. Tovmasyan, E. van Nieuwenburg, S. Huber.
\newblock {Phys. Rev. B} \textbf{88}, 220510(R) (2013).

\bibitem{Phillips2014} L.~G Phillips, G. {De Chiara}, P.Öhberg,
M. Valiente. \newblock {Phys. Rev. B} \textbf{91}, 054103 (2015).

\bibitem{Pudleiner2015} P. Pudleiner, A. Mielke. \newblock {Eur.
Phys. J. B} \textbf{88}, 207 (2015).

\bibitem{Gremaud2016} B~Grémaud, G~G Batrouni. \newblock {arXiv
preprint} (1612.00550) (2016).

\bibitem{Zhitomirsky2004} M.E. Zhitomirsky, H~Tsunetsugu. \newblock
{Phys. Rev. B} \textbf{70}, 100403(R) (2004).

\bibitem{Schmidt2006}H.-J. Schmidt, J. Richter, and R. Moessner.
J. Phys. A: Math. Gen. 39, 10673 (2006).

\bibitem{Motruk2012} J. Motruk, A. Mielke. \newblock {J. Phys.
A} \textbf{45}, 225206 (2012).

\bibitem{Bolobas79} B~Bollobás. \newblock {\em {Graph theory}}.
\newblock Springer Verlag Berlin, Heidelberg, New York, 1979.

\bibitem{Mielke1992a} A~Mielke. \newblock {J. Phys. A: Math. Gen.}
\textbf{25}, 4335–4345 (1992).

\bibitem{Azaria1998}P. Azaria, C. Hooley, P. Lecheminant, C. Lhuillier,
and A. M. Tsvelik. Phys. Rev. Lett. 81, 1694 (1998). 

\bibitem{Waldtmann2000}Ch. Waldtmann, H. Kreutzmann, U. Schollwöck,
K. Maisinger, and H.-U. Everts. Phys. Rev. B 62, 9472 (2000).

\bibitem{Taie2015} S.~Taie, H.~Ozawa, T.~Ichinose, T.~Nishio,
S.~Nakajima, Y.~Takahashi. \newblock {Sci. Adv.} \textbf{1},
e1500854 (2015).
\end{thebibliography}
\end{document}